\documentclass[11pt]{article}

\usepackage[top=1in, left=1.2in, right=1.2in, bottom=1.5in]{geometry}

\usepackage{amssymb,amsmath}
\usepackage{amsthm}
\usepackage{fancyhdr}
\usepackage{tikz}
\usetikzlibrary{decorations.markings}

\newcommand{\RR}{{\mathbb R}}
\newcommand{\CC}{{\mathbb C}}

\newcommand{\HH}{\mathcal{H}}

\DeclareMathOperator{\sech}{sech}
\DeclareMathOperator{\Tanh}{tanh}

\begin{document}

\title{$N$-soliton formula and blowup result of the Wadati-Konno-Ichikawa equation}

\author{Hsiao-Fan Liu and Yusuke Shimabukuro\\
{\small \it Institute of Mathematics, Academia Sinica, Taipei, Taiwan, 10617} \\
}

\date{\today}
\maketitle

\begin{abstract}
We formulate the $N$ soliton solution of the Wadati-Konno-Ichikawa equation that is determined by purely algebraic equations. Derivation is based on the matrix Riemann-Hilbert problem. We give examples of one soliton solution that include smooth soliton, bursting soliton, and loop type soliton. In addition, we give an explicit example for two soliton solution that blows up in a finite time.
\end{abstract}

\section{Introduction}
The Wadati-Konno-Ichikawa (WKI) equation is given by
\begin{equation} \label{WKI}
iq_t+\left(\frac{q}{\sqrt{1+|q|^2}}\right)_{xx}=0, 
\end{equation}
where $q(x,t)$ is a complex function on $(x,t)\in \RR\times [0,\infty)$. Two different types of the WKI equations are first introduced in \cite{Wadati-Konno-Ichikawa-1979}. The first type corresponds to \eqref{WKI}, which is viewed as a nonlinear Schr\"{o}dinger (NLS) type equation, since the linearized operator of \eqref{WKI} is the Schr\"{o}dinger operator. Physically it is relevant to modelling the vortex filament in the low temperature regime. Recently, a formal derivation of the WKI equation is given in \cite{Gorder-2016} that follows from the Hamiltonian equation for a vortex line originally derived in \cite{Svistunov-1995}. The WKI equation, being one of exotic soliton equations, admits a bursting soliton whose maximum hight is infinity. This was first shown by Shimizu and Wadati \cite{Shimizu-Wadati-1980} by the Gelfand-Levitan formulation to derive soliton solution. The construction relies on existence of a Lax operator whose eigenvalue appears as a nontrivial parameter of soliton solution. The WKI equation can be written in the form $\partial_{xt}\psi=\partial_{tx}\psi$ where $\psi$ is a solution to the Wadati-Konno-Ichikawa spectral problem, given by  
\begin{equation} \label{WKI-spectral}
\psi_x=[i\lambda \sigma_3-\lambda \mathcal{M}]\psi, \quad \mathcal{M}=\begin{pmatrix} 0 & q\\ -\bar{q} & 0 \end{pmatrix},
\end{equation} 
and
\begin{equation} \label{WKI-spectral-t}
\psi_t=[-2i\lambda^2\sigma_3-2i\frac{1-\sqrt{1+|q|^2}}{\sqrt{1+|q|^2}}\lambda^2\sigma_3+2\frac{1}{\sqrt{1+|q|^2}}\lambda^2 \mathcal{M}-i\lambda \; \mathcal{N}]\psi 
\end{equation}
where $ \mathcal{N}=\begin{pmatrix} 0 & \left(\frac{q}{\sqrt{1+|q|^2}}\right)_x\\ \left(\frac{\bar{q}}{\sqrt{1+|q|^2}}\right)_x & 0 \end{pmatrix}$ and $\sigma_3=\begin{pmatrix} 1 & 0\\ 0 &-1\end{pmatrix}.$\\
It is shown in \cite{Wadati-Sogo-1983, Ishimori-1982} that the WKI spectral problem is related to the Ablowitz-Kaup-Newell-Segur (AKNS) type spectral problem under gauge transformations. This idea can be adapted for investigating auto-B\"{a}cklund transformation, e.g., see \cite{Boiti-Gerdjikov-Pempinelli-1986, Kundu-1987, Levi-Ragnisco-Sym-1984}.
 


 Among other integrable equations associated with the WKI spectral problem, the Short-Pulse (SP) equation has been most extensively studied. 
The SP equation is related to the Sine-Gordon equation whose exact solutions are used to obtain exact solutions of the SP equation \cite{Sakovich-2006, Matsuno-2007}. Loop soliton and breather solutions of the SP equation have been studied in numerous literatures. 
Furthermore, a wave breaking result for the SP equation has been shown rigorously in \cite{Liu-Pelinovsky-Sakovich-2009}.    
On the other hand, there has been little result on the WKI equation \eqref{WKI}. From the fact that the WKI equation admits a bursting soliton, a naive question to ask is whether or not there exists a solution whose maximum hight blows up to infinity in a finite time.
Existence of such blowup solution is not the case of many other integrable evolution equations. This question is interesting in problems of well-posedness and stability of soliton. 

 Recently, global dispersive solution to the WKI equation has been studied by one of us in \cite{Shimabukuro-present}, that is, if an initial data is smooth and small enough, then a corresponding solution remains smooth for all time. The matrix Riemann-Hilbert problem for the WKI spectral problem was formulated rigorously. In the present paper, we derive a fully algebraic formula of $N$ soliton solutions by summarizing a rigorous approach \cite{Shimabukuro-present} of the forward scattering problem and the inverse scattering problem which we relates to the framework of the matrix Riemann-Hilbert problem. The given formula, unarguably easier, will enable for further investigation into soliton behavior for the WKI equation. At present, we give a finite-time blowup soliton solution, i.e., the maximum hight of a smooth initial profile goes to infinity in a finite time. A blowup soliton is constructed in the case of two soliton case by \emph{mixing} a smooth soliton and a bursting soliton with suitably chosen parameters. 

It is worth mentioning that the Gelfand-Levitan equation by Shimizu and Wadati \cite{Shimizu-Wadati-1980} could be extended to a $N$ soliton formula while an implicit equation (in our paper, it is expressed as an algebraic equation \eqref{eps}) would remain to be non-algebraic and cumbersome to solve. Furthermore, their construction requires a potential $q$ to have a compact support. In our approach, this assumption can be significantly relaxed (see \cite{Shimabukuro-present}). Additionally, we note that it is not clear how the B\"{a}cklund transformation approach for the WKI equation leads to explicit $N$ soliton formula, while an idea of taking use of relation between the WKI system and the ANKS system is clear. Hence the B\"{a}cklund transformation for the WKI equation requires further investigation. 

It is well-known that the isotropic Heisenberg equation is related to the cubic NLS \cite{Zakharov-Takhtadzhyan-1978, Zakharov-Mikhailov-1978}. This relation is written explicitly in equation (4.7) of \cite{Wadati-Sogo-1983}.
 As mentioned previously, formally solution of the WKI equation is also related to the isotropic Heisenberg equation, explicitly written in equation (4.9) of \cite{Wadati-Sogo-1983} by Wadati and Sogo. A simple and interesting question is, how does the WKI equation allow bursting solitons, while the isotropic Heisenberg equation and the cubic NLS are known not to have such explicit solutions? Unlike relation between the cubic NLS and the isotropic Heisenberg equation, solution of the WKI equation contains the inverse power of that of the isotropic Heisenberg equation. This latter relation is not well-defined. A singularity formation for the WKI equation appears to be quite possible.

 
The paper is organized as follows.

 In Section \ref{soliton-derivation}, we give the algebraic formulas that give explicit solutions to the WKI equation.

  
 In Section \ref{Riemann-Hilbert-section}, we summarize part of \cite{Shimabukuro-present} to give derivation of formulas introduced in Section \ref{soliton-derivation}. there are mainly three important steps for technical aspects, (i) transformation of the WKI spectral problem to the ANKS spectral problem for existence and analyticity of the fundamental solutions, (ii) the change of spectral parameter $\lambda \rightarrow -\frac{1}{\lambda}$ for suitable normalization, (iii) the change of the spatial variable in \eqref{y} that depends on the dependent variable to eliminate the potential $q$ in the jump condition of the Riemann-Hilbert problem. 
This type of transformation in (iii) is called the reciprocal transformation or the hodograph transformation in many literatures, e.g., see \cite{Feng-Inoguchi-Kajiwara-Maruno-2011} for integrable discretizations. 

In Sections \ref{one-soliton-example}, we give three explicit examples of one soliton solutions, which are smooth soliton, bursting soliton, loop type soliton. 

In Section \ref{two-soliton-example}, we consider two soliton solution with fixed spectral parameters which correspond to a smooth soliton and a bursting soliton, if they are considered individually in Section \ref{one-soliton-example}. We give two different examples. One is that two soliton solution consists simply of smooth soliton and bursting soliton from initial time, as expected. The other is that the maximum hight of smooth two soliton at initial time blows up to infinity in a finite time. This example is shown in Subsection \ref{blowup-section}.  

\section{$N$-soliton formula} \label{soliton-derivation}
We denote $q_N(x,t)$ as $N$-soliton solution of the WKI equation. In order to derive $q_{N}(x,t)$, we will first obtain the intermediate solution, which is denoted as $\tilde{q}_{N}(y,t)$ that depends on the spatial variable $y$.   

We denote sets of parameters $\{\beta_1, \cdots, \beta_N\} \subset \CC$ and $\{\lambda_1,\cdots, \lambda_N\} \subset \CC^+$, where $\CC^+$ is the upper-half plane of the complex plane $\CC$, and $N$ corresponds to the subscript $N$ of $q_N$. We define $M_{ij}$ by
$$M_{ij}:=\frac{\beta_i}{\lambda_i^{-1}-\bar{\lambda}_j^{-1}},$$
and we define $\theta(\rho)$ by
$$\theta(\rho):=-2i\rho y+4i\rho^2 t.$$
For example, we have that $\theta(\lambda_i)=-2i\lambda_i y+4i\lambda_i^2 t$. We shall build $N \times N$ matrices $A$ and $B$ using above notations,
\begin{equation} \label{A-B}
A=[A_{k \ell}], \quad B=[B_{k \ell}],
\end{equation}
whose elements at the $k$ th row and the $\ell$ th column are given by
$$A_{k\ell}=\sum_{j=1}^NM_{jk}\overline{M}_{\ell j}e^{\theta(\lambda_j)-\theta(\bar{\lambda}_{\ell})},  \quad B_{k\ell}=\sum_{j=1}^N\overline{M}_{jk}M_{\ell j}e^{\theta(\lambda_{\ell})-\theta(\bar{\lambda}_{j})}.$$ 
We see that $B_{k\ell}=\overline{A}_{k\ell}$. Lastly, we define a $N\times 1$ vector $\mathbf{b}$,
\begin{equation} \label{b}
\mathbf{b}=[b_k], \quad b_k=\sum_{j=1}^N\overline{M}_{jk}e^{-\theta(\bar{\lambda}_{j})}
\end{equation}
where $b_k$ is the $k$ th element in the vector.

 Using the above notations, we introduce systems of algebraic equations for vector functions $\alpha=(\alpha_1,\cdots,\alpha_N)^t$ and $\gamma=(\gamma_1,\cdots,\gamma_N)^t$,
\begin{equation} \label{alpha-system}
(I+A)\alpha = \mathbf{1}, \quad (I+B)\gamma=\mathbf{b},
\end{equation}
where $I$ is a $N\times N$ identity matrix, $\mathbf{1}$ is a $N\times 1$ vector with one in each entry.  

The intermediate $N$-soliton solution $\tilde{q}_{N}(y,t)$, defined on the variable $y$, is found by the following formula
\begin{equation} \label{N-soliton-y}
\tilde{q}_{N}(y,t)=\frac{1}{\sqrt{1-|\partial_yg|^2}}\partial_yg, \quad g(y,t)=\sum_{j=1}^N\bar{\beta}_j e^{-\theta(\bar{\lambda}_j)}\alpha_j(y,t)
\end{equation}
where $\alpha_j$'s are found in \eqref{alpha-system}. The original variable $x$ is found from
\begin{equation} \label{x-equation}
y=x+\epsilon(x,t),
\end{equation}
and $\epsilon$ is determined from the implicit function
\begin{equation} \label{eps}
\epsilon(x,t)=f(x+\epsilon(x,t),t), \quad f(y,t)= i\sum_{j=1}^N\beta_j e^{-\theta(\lambda_j)}\gamma_j(y,t),
\end{equation}
where $\gamma_j's$ are found in \eqref{alpha-system}.

Finally, $N$-soliton solution $q_{N}$ is obtained by \eqref{N-soliton-y} and \eqref{x-equation},
\begin{equation} \label{q-N}
q_N(x,t)=\tilde{q}_N(x+\epsilon(x,t),t)
\end{equation}
where $\epsilon(x,t)$ satisfies the implicit function \eqref{eps}.

To summarize the above presentation, initially we fix $\{\beta_j,\lambda_j\}_{j=1}^N$ and follow the steps, given as  
$$\{\beta_j,\lambda_j\}_{j=1}^N \longrightarrow \tilde{q}_N(y,t)\;\; \mbox{from}\; \eqref{N-soliton-y} 
 \longrightarrow \tilde{q}_N(x+\epsilon(x,t),t),$$
 where
 $$ \epsilon(x,t) = f(x+\epsilon(x,t),t) \;\; \mbox{from}\; \eqref{eps}.$$






\section{Derivation of $N$-soliton formula} \label{Riemann-Hilbert-section}
Here, we summarize the results from \cite{Shimabukuro-present}. (There are a few notational changes for convenience.) 
\subsection{Transformation to the AKNS type spectral problem}
We shall introduce the normalized solution $m^{(\pm)}=\psi^{(\pm)}e^{-i\lambda \sigma_3x}$ to the WKI spectral problem \eqref{WKI-spectral} and write the integral systems from \eqref{WKI-spectral}, 
\begin{equation}\label{m-integral}
m^{(\pm)}=I-\lambda\int_{\pm\infty}^x e^{i\lambda(x-x')\sigma_3}\mathcal{M}m^{(\pm)}e^{-i\lambda(x-x')\sigma_3}dx'.
\end{equation}
We use $(\pm)$ above to mean that $m^{(\pm)}$ is the normalized solution at $x=\pm \infty$. Instead of working with the integral equation \eqref{m-integral} directly, it is convenient to carry out the following transformations in the WKI spectral problem \eqref{WKI-spectral},
\begin{equation}\label{psi-varphi}
\psi^{(\pm)}=\mathcal{G}g_{\pm \infty} \psi_{\mbox{\tiny AKNS}}^{(\pm)}
\end{equation}
where
\begin{align*}
\mathcal{G} &=\frac{1}{\sqrt{2}(1+|q|^2+\sqrt{1+|q|^2})^{1/2}}\begin{pmatrix}  1+\sqrt{1+|q|^2} & -iq \\ -i\bar{q} & 1+\sqrt{1+|q|^2} \end{pmatrix}, \\
g_{\pm\infty}&=\exp\left( \sigma_3 \int_{\pm \infty}^xB(x')dx'\right)\; \mbox{with}\; B= \frac{1}{4}\frac{q_x\bar{q}-q\bar{q}_x}{1+|q|^2+\sqrt{1+|q|^2}}.
\end{align*}
We note that $\det(\mathcal{G})=1$ and $|g_{\pm\infty}|=1$ since $B$ is purely imaginary.
We find that $\psi_{\mbox{\tiny AKNS}}^{(\pm)}$ satisfy \emph{the ANKS type spectral problem}, given as 
\begin{equation}\label{AKNS}
\partial_x \psi_{\mbox{\tiny AKNS}}^{(\pm)}=[\sigma_3 i\lambda\sqrt{1+|q|^2} + \mathcal{V}] \psi_{\mbox{\tiny AKNS}}^{(\pm)}, \quad \mathcal{V}= g_{\pm\infty}^{-1}\begin{pmatrix}0 & Q \\ -\bar{Q} & 0 \end{pmatrix}g_{\pm\infty},
\end{equation}
where 
$$Q= \frac{-i}{4(1+|q|^2+\sqrt{1+|q|^2})}(q(\sqrt{1+|q|^2})_x -q_x(1+\sqrt{1+|q|^2})).$$

\subsection{Analytic properties of $m^{(\pm)}(\lambda)$ and scattering coefficients}
We shall introduce the normalized solution $\varphi^{(\pm)}$ by
$$ \varphi^{(\pm)} =  \psi_{\mbox{\tiny AKNS}}^{(\pm)} e^{-i\lambda \sigma_3p_{\pm\infty}(x)},$$ 
where
$p_{\pm\infty}(x)=x + \int_{\pm\infty}^x(\sqrt{1+|q|^2}-1)dy.$ 
Analytic properties of $\varphi^{(\pm)}$ in the $\lambda$ variable can be rigorously studied if $\int_{\RR}|Q|dx<\infty$ for the ANKS type spectral problem \eqref{AKNS} (e.g, see Lemma 2.1 in \cite{Ablowitz-Prinari-Trubatch-2004}). Analyticities of $m^{(\pm)}(\lambda)$ are determined by those of $\varphi^{(\pm)}(\lambda)$ since $\det(\mathcal{G})=1$, $B$ is purely imaginary, and $\mathcal{G}, B$ are independent of $\lambda$. 

More precisely, we have
$$[(\varphi^{(+)})_1, (\varphi^{(-)})_2] \;\mbox{is analytic in} \; \CC^+, \quad [(\varphi^{(-)})_1, (\varphi^{(+)})_2] \;\mbox{is analytic in} \; \CC^-,$$ 
where we denote $(\varphi^{(\pm)})_{j}$ as the $j$ th column of matrix $\varphi^{(\pm)}$, and $\CC^{\pm}$ as the upper/lower half planes of the $\lambda$-complex plane.

Furthermore, as $|\lambda|\rightarrow \infty$ in their domains of analyticity, we have 
\begin{equation}\label{asymptotic}
[(\varphi^{(+)})_1, (\varphi^{(-)})_2] \rightarrow I, \quad [(\varphi^{(-)})_1, (\varphi^{(+)})_2] \rightarrow I.
\end{equation} 

From \eqref{psi-varphi}, it follows that columns of $m^{(\pm)}$ shares the same analyticities as those of $\varphi^{(\pm)}$ with the limits,
\begin{equation} \label{m-lambda-infinity-plus}
\left\{\begin{matrix}
[m^{(+)}]_1e^{i\lambda \int_x^{\infty}(\sqrt{1+|q|^2}-1)dx'} &\rightarrow (\mathcal{G})_1 e^{-\int_x^{\infty}Bdx'}\\
[m^{(-)}]_2e^{i\lambda \int_{-\infty}^x(\sqrt{1+|q|^2}-1)dx'} &\rightarrow (\mathcal{G})_2 e^{-\int_{-\infty}^xBdx'}
\end{matrix}\right. \;\mbox{as}\; \CC^+\ni \lambda \rightarrow \infty,
\end{equation}
and 
\begin{equation} \label{m-lambda-infinity-minus}
\left\{\begin{matrix}
[m^{(-)}]_1e^{-i\lambda \int_{-\infty}^x(\sqrt{1+|q|^2}-1)dx'} &\rightarrow (\mathcal{G})_1 e^{\int_{-\infty}^xBdx'}\\
[m^{(+)}]_2e^{-i\lambda \int^{\infty}_x(\sqrt{1+|q|^2}-1)dx'} &\rightarrow (\mathcal{G})_2 e^{\int^{\infty}_xBdx'}
\end{matrix}\right. \;\mbox{as}\; \CC^-\ni \lambda \rightarrow \infty.
\end{equation}
 We define the scattering coefficients $a$ and $b$ by relations between $\psi^{(+)}$ and $\psi^{(-)}$,
\begin{equation} \label{def-T}
\psi^{(+)}=\psi^{(-)}T, \quad T =\begin{pmatrix} a(\lambda) & -\overline{b(\bar{\lambda})} \\ b(\lambda) & \overline{a(\bar{\lambda})} \end{pmatrix}, 
\end{equation}
where $\bar{a}$ is a complex conjugate of $a$.
We see that $a(\lambda)$ can be written as the determinant 
\begin{equation}\label{a-det}
a(\lambda)=\det((\psi^{(+)})_1,(\psi^{(-)})_2)=\det((m^{(+)})_1,(m^{(-)})_2)
\end{equation}
which implies that $a(\lambda)$ is analytic in $\CC^+$ since $(m^{(+)})_1$ and $(m^{(-)})_2$ are analytic in $\CC^+$.

The limit of $a(\lambda)$ as $|\lambda|\rightarrow \infty$ can be seen from \eqref{m-lambda-infinity-plus} and \eqref{a-det}, so it follows that $a(\lambda)$ is analytic in $\CC^+$ with the limit
\begin{equation}\label{a-limit}
a(\lambda)e^{i\lambda \int_{\RR} (\sqrt{1+|q|^2}-1)dx'} \rightarrow e^{- \int_{\RR}Bdx'}, \quad \CC^+\ni \lambda \rightarrow \infty.
\end{equation}

\subsection{The change in the spectral parameter $z=-\frac{1}{\lambda}$}
From \eqref{m-lambda-infinity-plus} and \eqref{m-lambda-infinity-minus}, we see that the matrix functions $[(m^{(\pm)})_1,(m^{(\mp)})_2]$ are not normalized to identity at $|\lambda|=\infty$, unlike $[(\varphi^{(\pm)})_1,(\varphi^{(\mp)})_2]$ that tend to $I$ as $|\lambda|\rightarrow \infty$ in their domains of analyticities. For this reason, we introduce the new spectral parameter 
$$z=-\frac{1}{\lambda}.$$ 
We easily see that $\mbox{Im}(z)>0$ if and only if $\mbox{Im}(\lambda)>0$, which implies that 
 $$[(m^{(+)}(z))_1, (m^{(-)}(z))_2] \;\mbox{is analytic in} \; \CC^+_z, \quad [(m^{(-)}(z))_1, (m^{(+)}(z))_2] \;\mbox{is analytic in} \; \CC^-_z,$$ 
where we use $\CC_z$ to denote the $z$-complex plane. From the integral equation \eqref{m-integral} by introducing $z=-1/z$,  
  \begin{equation}\label{asymptotic}
[(m^{(+)}(z))_1, (m^{(-)}(z))_2] \rightarrow I, \quad [(m^{(-)}(z))_1, (m^{(+)}(z))_2] \rightarrow I.
\end{equation}
 as $|z|\rightarrow \infty$ in their domains of analyticity. 
Furthermore, the integral equation \eqref{m-integral} gives that the potential $\mathcal{M}$ is recovered from the derivative of the first term $m_1$ in the series
\begin{equation}\label{asymptotic-2}
[(m^{(+)})_1, (m^{(-)})_2]=I+\frac{m_1}{z} + \mathcal{O}(\frac{1}{z^2}),
\end{equation}
that is,
\begin{equation}\label{potential-formula}
\partial_xm_1=\mathcal{M}=\begin{pmatrix} 0 & q\\ -\bar{q} & 0 \end{pmatrix}.
\end{equation}

Similarly, for $a$ in \eqref{a-det} on the $z$ variable, we have that $a(-\frac{1}{z})$ is analytic in $\CC_z^+$, and the determinant form of $a$ in \eqref{a-det} with the above asymptotic behavior \eqref{asymptotic-2} gives
\begin{equation}\label{a-z-limit}
a(-\frac{1}{z})=1+\mathcal{O}(\frac{1}{z^2}),
\end{equation}
for $z\in \CC^+_z$ in the neighborhood of $\infty$. 
To this end, we introduce the following sectionally meromorphic matrix function
\begin{equation} \label{normalized-m}
m(z)=\left\{\begin{matrix} m_+(z) & z \in \CC^+ \\
 m_-(z) & z \in \CC^- 
 \end{matrix} \right.
 \end{equation}
with (denote $\HH=(\sqrt{1+|q|^2}-1)$)
 \begin{align*}
 m_+(z)&= \left[ \frac{1}{a(-1/z)e^{-\frac{i}{z}\int_{\RR}\mathcal{H}dy}} (m^{(+)})_1e^{-\frac{i}{z}\int_{x}^{\infty}\mathcal{H}dy}, (m^{(-)})_2e^{-\frac{i}{z}\int_{-\infty}^x\mathcal{H}dy}\right],\\ m_-(z)&= \left[(m^{(-)})_1e^{\frac{i}{z}\int_{-\infty}^x\mathcal{H}dy},  \frac{1}{\overline{a}(-1/\bar{z})e^{\frac{i}{z}\int_{\RR}\mathcal{H}dy}} (m^{(+)})_2e^{\frac{i}{z}\int_{x}^{\infty}\mathcal{H}dy}\right].
 \end{align*}
 From the previous discussion, we see that $m_{\pm}(z)$ are analytic in $\CC^{\pm}_z$ except at the possible zeros of $a$ and $\overline{a}$. The matrix function $m(z)$ satisfy the following limits, from \eqref{asymptotic} and \eqref{a-z-limit}, 
 \begin{equation}\label{m-limit-1}
m_{\pm}(z)\rightarrow I, \quad |z|\rightarrow \infty \quad (z \in \CC^{\pm})
\end{equation}
and from \eqref{m-lambda-infinity-plus}, \eqref{m-lambda-infinity-minus}, and \eqref{a-limit},
\begin{equation}\label{m-limit-2}
m_{\pm}(z) \rightarrow \mathcal{G}e^{\sigma_3\int_{-\infty}^xBdy} \quad |z|\rightarrow 0 \quad (z \in \CC^{\pm}).
 \end{equation}
Furthermore, from \eqref{def-T}, we can directly compute and verify that 
 \begin{equation} \label{jump-RHP}
m_+(z)= m_-(z) \begin{pmatrix} 1+|r|^2 & \bar{r}e^{-\frac{2i}{z}(x+\int_{-\infty}^x\mathcal{H}dx')} \\
r e^{\frac{2i}{z}(x+\int_{-\infty}^x\mathcal{H}dx')} & 1 \end{pmatrix} \quad z\in \RR,
\end{equation}
 where we introduced the reflection coefficient $r(z):=\frac{b(-1/z)}{a(-1/z)}$. The condition \eqref{jump-RHP} is the jump condition for $m(z)$ in \eqref{normalized-m} on the real line. Smoothness and decay rate of $r(-1/\lambda)$ depends on these of the potential $q(x)$. If $q$ has sufficient smoothness and decay rate, then $r$ is bounded and $r(\pm\infty)=0$, and $r(0)=0$. It is important to note that at $z=0$, \eqref{jump-RHP} is 
 $$m_{+}(0)=m_-(0),$$   
 which coincides with the limit \eqref{m-limit-2}. This indicates the correct normalization in the form of $m(z)$ defined in \eqref{normalized-m}.
 
From \eqref{m-limit-1}, \eqref{jump-RHP}, and the fact that $m_{\pm}(z)$ is analytic in $\CC^{\pm}_z$ except at zeros of $a$, the matrix function $m(z)$ in \eqref{normalized-m} can be viewed as the solution of the normalized matrix Riemann-Hilbert (RH) problem. We have related the WKI spectral problem to the RH problem. 
 
From $m(z)$, we shall give two important formulas that recover the potential $q$. We denote $m_{ij}$ as the $(i,j)$ th element of the matrix $m(z)$. We see that, from \eqref{potential-formula}, one of important formulas is
\begin{equation}\label{recover-q} 
q=\partial_x(m_{12}^{(1)}), \quad m_{12}^{(1)}:=\lim_{|z|\rightarrow \infty}z\;m_{12}(z).  
\end{equation}
From \eqref{asymptotic-2}, \eqref{a-z-limit}, and definition of \eqref{normalized-m}, we have 
 $$m_{11}(z)=1+\frac{i}{z}\int_{-\infty}^{x}(\sqrt{1+|q|^2} -1)dx'+\mathcal{O}(z^{-2}),$$
for $z\in \CC^+_z$ in the neighborhood of $\infty$, so the second important formula is given as 
\begin{equation}\label{recover-H}
i\int_{-\infty}^{x}(\sqrt{1+|q|^2} -1)dx'=m_{11}^{(1)}, \quad m_{11}^{(1)}:=\lim_{|z|\rightarrow \infty}z\; m_{11}(z).
\end{equation}  
 
\subsection{Time evolution of $r(z)$} Up to this moment, we have not considered the time evolution of the WKI equation. Here we shall give a brief argument for the time evolution of reflection coefficient $r$ under the WKI time evolution $q(x,t)$.
 
Let $q(x,t)$ be solution to the WKI equation. Taking $|x|\rightarrow \infty$ in the Lax system \eqref{WKI-spectral-t} with $q(x,t)$, we obtain $\psi_t=-2i\lambda^2\sigma_3\psi$. This implies that the fundamental solutions that satisfy the both Lax systems must take form of $\psi^{(\pm)}(x,t;\lambda)e^{-2i\lambda^2t\sigma_3}$. Following the exactly same definition of the scattering coefficients $a$ and $b$ with $\psi_{\pm}(x,t,;\lambda)$ in \eqref{def-T}, we find 
\begin{equation} \label{a-b-time}
a(\lambda,t)=a(\lambda), \quad b(\lambda,t)=e^{4i\lambda^2t}b(\lambda).
\end{equation}
We find that the reflection coefficient $r$ depends on $t$ as $r(z,t)=r(z)e^{4i\frac{t}{z^2}}.$ What follows is that we will simply replace $r$ in \eqref{jump-RHP} by $r(z)e^{4i\frac{t}{z^2}}$ to take the time evolution into account. 
 
\subsection{The change in the spatial coordinate} 
If we want to derive exact solutions from the RH formulation, we do not have any knowledge of $q$ that appears in the jump condition \eqref{jump-RHP}. In order to address this issue, we introduce the change of the spatial coordinate 
\begin{equation}\label{y}
y:=x+\int_{-\infty}^x(\sqrt{1+|q|^2}-1)dx', \quad \frac{dy}{dx}=\sqrt{1+|q|^2}.
\end{equation}
We shall denote $\tilde{q}(y,t)$ that is related to $q(x,t)$ as follows, 
\begin{equation} \label{q-tilde}
\tilde{q}(y,t)=\tilde{q}\left(x+\int_{-\infty}^x(\sqrt{1+|q|^2}-1)dx',t\right)=q(x,t).
\end{equation}
We apply the above change of variable to the reconstruction formula \eqref{recover-q} and obtain,
$$\partial_ym_{12}^{(1)}=\frac{\tilde{q}(y,t)}{\sqrt{1+|\tilde{q}(y,t)|^2}}.$$
From the above equation, we solve for $\tilde{q}(y)$ to obtain
\begin{equation} \label{recover-q-2}
\tilde{q}(y,t)=\frac{\partial_ym_{12}^{(1)}}{\sqrt{1-\left|\partial_ym_{12}^{(1)}\right|^2}},
 \end{equation}
where $m_{12}$ is obtained from the solution $m(z;y)$ to the normalized RH problem, i.e., given $r(z) \in \RR$, find sectionally meromorphic functions $m(z;y)$ in $\CC^{\pm}$ such that 
\begin{equation} \label{RHP-jump-2}
 m_+(z)= m_-(z) \begin{pmatrix} 1+|r|^2 & \bar{r}e^{-\frac{2i}{z}y-4i\frac{t}{z^2}} \\
r e^{\frac{2i}{z}y+4i\frac{t}{z^2}} & 1 \end{pmatrix} \quad z\in \RR,
\end{equation}
and $m(z)\rightarrow I$ as $|z|\rightarrow \infty$ in $\CC^{\pm}$. We see that $\tilde{q}(y,t)$ in \eqref{recover-q-2} is determined from solution $m(z;y)$ to the RH problem without any knowledge of $q$. The equation \eqref{recover-q-2} in fact corresponds to \eqref{N-soliton-y}.
By denoting 
$$\epsilon(x,t):=\int_{-\infty}^x(\sqrt{1+|q|^2}-1)dx'$$
and 
by definitions of the $y$ variable in \eqref{y} and of $\tilde{q}(y,t)$ in \eqref{q-tilde}, we have 
\begin{equation} \label{eps-def}
y=x+\epsilon(x,t), \quad \epsilon(x,t)=\int_{-\infty}^x(\sqrt{1+|\tilde{q}(x'+\epsilon(x',t),t)|^2}-1)dx'.
\end{equation}
We can avoid the integral expression above by noticing that, from \eqref{recover-H}, 
\begin{equation} \label{eps-recover}
\epsilon(x,t)=\frac{1}{i}m_{11}^{(1)}(x+\epsilon(x,t),t).
\end{equation}
The argument of $m_{11}^{(1)}$ is written as $x+\epsilon(x,t)$ since $m_{11}^{(1)}$ is determined on the $y$ variable. We will see that $\frac{1}{i}m_{11}^{(1)}$ in fact is related to $f$ in \eqref{eps}.

\subsection{Derivations of \eqref{alpha-system}, \eqref{N-soliton-y}, and \eqref{eps}}
 For convenience, we express components of $m(z)$ in \eqref{normalized-m} as 
$$f^+:= (m^{(+)})_1e^{-\frac{i}{z}\int_x^{\infty}\HH dy}, \quad g^+:= (m^{(+)})_2e^{-\frac{i}{z}\int_{-\infty}^x\HH dy},$$
$$f^-:= (m^{(-)})_1e^{\frac{i}{z}\int_{-\infty}^x\HH dy}, \quad g^-:= (m^{(-)})_2e^{\frac{i}{z}\int_x^{\infty}\HH dy}.$$
and 
$$c(z):=a(-1/z)e^{-\frac{i}{z}\int_x^{\infty}\HH dy}.$$
We simply have the expression
$$m(z) = \left\{ \begin{matrix}m_+(z)= \left[ \frac{1}{c(z)}f^+, g^+ \right] & z \in \CC^+ \\
 m_-(z)=[f^-,  \frac{1}{\overline{c(\bar{z})}}g^-] & z \in \CC^- 
 \end{matrix} \right.
$$
We make the following assumptions on $a(-1/z)$ and $r(z)$:\\
(I) $a(-1/z)$ has $N$ simple zeros at $z=z_j \in \CC^+$ ($j=1,\cdots, N$)\\
(II) $r(z)=0$\\ 
The goal is to find the corresponding potential $q$ under above assumptions.
The assumption (II) implies that, from \eqref{RHP-jump-2}, $m(z)$ satisfies 
\begin{equation}\label{jump-r-zero}
m_+(z)=m_-(z), \quad z\in \RR.
\end{equation}
The assumption (I) implies that, from $a=\det((\psi^{(+)})_1,(\psi^{(-)})_2)=0$ at $z=z_{j}$, $(\psi^{(+)})_1$ and $(\psi^{(-)})_2$ are linearly dependent at $z=z_{j}$. This yields the following relation 
\begin{equation} \label{relation-1}
f^+(z_j)=k_j e^{\frac{2i}{z_j}y+\frac{4i}{z_j^2}t}g^+(z_j), \quad j=1,2,\cdots, N,
\end{equation}
where $k_j$ is some constant. Furthermore, by symmetry of the WKI spectral problem, it is easy to verify that $g^-(\bar{z})=\begin{pmatrix} 0 & -1\\ 1 &0\end{pmatrix} \overline{f^+(z)}$ and $f^-(\bar{z})=\begin{pmatrix} 0 & 1\\ -1 &0\end{pmatrix} \overline{g^+(z)}$. Applying these relations to \eqref{relation-1}, we have
\begin{equation} \label{relation-2}
g^-(\bar{z}_j)=-\bar{k}_j e^{-\frac{2i}{\bar{z}_j}y-\frac{4i}{\bar{z}_j^2}t}f^-(\bar{z}_j),  \quad j=1,2,\cdots, N.
\end{equation}

Let the contour $C^+ \subset \CC^+$ be union of a semi-circle and small circles that enclose each $z_j$, $j=1,2,\cdots,N$, as shown below
\begin{figure}[htbp] 
   \centering
 \begin{tikzpicture}[decoration={markings,
mark=at position 0.5 with {\arrow[line width=1pt]{>}},
}
]

\path[draw,postaction=decorate] (-3,0) -- (3,0);
\path[draw, postaction=decorate] (3,0) arc(0:180:3);

\path[draw, postaction=decorate] (-1,1) arc(360:0:0.2);
\path[draw, postaction=decorate] (-2,1) arc(360:0:0.2);
\path[draw, postaction=decorate] (1,1) arc(360:0:0.2);
\path[draw, postaction=decorate] (2,1) arc(360:0:0.2);

\path[draw] (-1.2,0) -- (-1.2,0.8) [dashed];
\path[draw] (-2.2,0) -- (-2.2,0.8) [dashed];
\path[draw] (0.8,0) -- (0.8,0.8) [dashed];
\path[draw] (1.8,0) -- (1.8,0.8) [dashed];

\draw[fill] (-1.2,1) circle [radius = 0.02];
\draw[fill] (-2.2,1) circle [radius = 0.02];
\draw[fill] (0.8,1) circle [radius = 0.02];
\draw[fill] (1.8,1) circle [radius = 0.02];


\node[above] at (0,1) {$D^+$};
\end{tikzpicture}
\end{figure}

Let $C^-\subset \CC^-$ be the contour that is the reflected contour of $C^+$ with respect to the $x$ axis after reversing orientations. Small clock-wise circles of $C^-$ enclose each $\bar{z}_j$, $j=1,2,\cdots,N$. The Cauchy integrable formula yields
$$\frac{1}{2\pi i}\int_{C^{\pm}}\frac{m(s)-I}{s-z}ds=\left\{ \begin{matrix} m(z)-I & z \in D^{\pm}
\\ 0 & z \in \CC^{\mp} \end{matrix}\right.
$$
We can express 
\begin{equation} \label{cauchy-integral}
m(z)-I=\frac{1}{2\pi i}\left(\int_{C^{+}}+\int_{C^{-}}\right)\frac{m(s)-I}{s-z}ds, \quad z\in D^+\cup D^-.
\end{equation}
In the above equation \eqref{cauchy-integral}, we carry out the following procedures--take the radii of the semi-circles in $C^{\pm}$ to infinity and of the small circles in $C^{\pm}$ to zero, the horizontal lines in $C^{\pm}$ to the real line, and apply relations \eqref{jump-r-zero}, \eqref{relation-1}, and \eqref{relation-2}. The resulting equation of equation \eqref{cauchy-integral} is  
\begin{equation} \label{m-residues}
m(z)=I-\sum_{j=1}^N\frac{\bar{\beta}_j e^{-\frac{2i}{\bar{z}_j}y-\frac{4i}{\bar{z}_j^2}t}}{\bar{z}_j-z}m(\bar{z}_j)\begin{pmatrix} 0 & 1 \\ 0 & 0\end{pmatrix}+\sum_{j=1}^N\frac{\beta_j e^{\frac{2i}{z_j}y+\frac{4i}{z_j^2}t}}{z_j-z}m(z_j)\begin{pmatrix} 0 & 0 \\ 1 & 0\end{pmatrix},
\end{equation}
where $\beta_j=-k_j/a'(z_j)$. 
We shall denote $\alpha_j:=(f^-(\overline{z}_j))_1$ and $\gamma_j:=(g^+(z_j))_1$ which are the first components of $f^-(\overline{z}_j)$ and $g^+(z_j)$. Algebraic equations for $\alpha_j$ and $\gamma_j$ are obtained from \eqref{m-residues} by setting $z=z_k$ for $(g^+(z))_1$ and $z=\bar{z}_j$ for $(f^-(z))_1$, $k=1,2,\cdots,N$. Thus, we obtain \eqref{alpha-system}.

From \eqref{m-residues}, we obtain
$$g(y,t):=m_{12}^{(1)}=\lim_{|z|\rightarrow}z\;(m(z))_{12}=\sum_{j=1}^N\bar{\beta}_j e^{-\frac{2i}{\bar{z}_j}y-\frac{4i}{\bar{z}_j^2}t} \alpha_j(y,t),$$
which gives the second equation in \eqref{N-soliton-y} after $z_j=-\frac{1}{\lambda_j}$. The first equation in \eqref{N-soliton-y} is already found in \eqref{recover-q-2}. 
 
 Similarly, from \eqref{m-residues}, we obtain  
 $$\frac{1}{i}\lim_{|z|\rightarrow \infty} z\; (m(z))_{11}=i\sum_{j=1}^N\beta_j e^{\frac{2i}{z_j}y+\frac{4i}{z_j^2}t}\gamma_j(y,t).$$ 
 The above equation is the explicit expression of $\epsilon(x,t)$ in \eqref{eps-recover}.
 We shall denote $f(y,t):= i\sum_{j=1}^N\beta_j e^{\frac{2i}{z_j}y+\frac{4i}{z_j^2}t}\gamma_j(y,t)$, which corresponds to \eqref{eps}.   
  
\section{One soliton solution $N=1$} \label{one-soliton-example}
In the following presentation, we choose $N=1$, $\beta_1=|\lambda_1^{-1}-\overline{\lambda}_1^{-1}|$, and we denote
$$\lambda_1=\xi+i\eta, \quad \xi, \eta \in \RR.$$ 
Solving for $\alpha_1$ in \eqref{alpha-system} and $g$ in \eqref{N-soliton-y}, we find that  
$$\partial_{y}g= \frac{4\eta}{\xi^2+\eta^2}e^{-i(-2\xi x_{\HH}+4(\xi^2-\eta^2))t} [\eta \Tanh(2\eta y-8\xi\eta t)-i\xi]\sech(2\eta y-8\xi\eta t),$$
and
\begin{equation}\label{q1-mag}
|\tilde{q}_{1}(y,t)|^2=\frac{|\partial_yg|^2}{1-|\partial_yg|^2}=\frac{4\eta^2}{\xi^2+\eta^2}\frac{\cosh^2(2\eta y-8\xi\eta t)-\frac{\eta^2}{\xi^2+\eta^2}}{[\cosh^2(2\eta y-8\xi\eta t)-\frac{2\eta^2}{\xi^2+\eta^2}]^2}.
\end{equation}
In order to solve for the original variable $x$, we solve for $\gamma_1$ in \eqref{alpha-system} and find that from \eqref{eps}
\begin{equation} \label{epsilon}
\epsilon(x,t)=\frac{\eta}{\xi^2+\eta^2}\left[\tanh(2\eta(x-4\xi t+\epsilon(x,t)))+1\right].
\end{equation}
One soliton solution is given by
\begin{align}
q_1(x,t) &= \tilde{q}_1(x+\epsilon(x,t),t) \nonumber\\
\label{one-soliton} &=2i\frac{\eta}{\sqrt{\xi^2+\eta^2}}e^{-i(-2\xi (x+\epsilon(x,t))+4(\xi^2-\eta^2)t)} \frac{ \cosh(2\eta (x+\epsilon(x,t))-8\xi\eta t+i\alpha)}{\left|\cosh^2(2\eta (x+\epsilon(x,t))-8\xi\eta t)-\frac{2\eta^2}{\xi^2+\eta^2}\right|},
\end{align}
where $\alpha= \arctan(\eta/\xi)$ and $\epsilon(x,t)$ is found by an implicit equation \eqref{epsilon}.
 This coincides with one soliton in the original paper \cite{Shimizu-Wadati-1980} after the change of parameter $\lambda \rightarrow -\lambda$ because of our Lax pair with opposite sign for $\lambda$. 
 
For this moment, it is convenient to use the variable $y=x+\epsilon(x,t)$. The denominator is denoted as 
$$D(y,t):=\cosh^2(2\eta y-8\xi\eta t)-\frac{2\eta^2}{\xi^2+\eta^2}.$$
We see that
$$D(y,t) >0\quad \mbox{if}\; |\xi|>|\eta|.$$
The case $|\xi|>|\eta|$ corresponds to smooth soliton solution.

On the other hand, it is easy to verify that, if $|\xi| \leq |\eta|$,
$$D(y^*,t)=0$$
when
\begin{equation}\label{zero-y}
y^*=4\xi t +\frac{1}{2\eta}\left[\ln\left(\sqrt{\frac{2\eta^2}{\xi^2+\eta^2}}\pm \sqrt{\frac{\eta^2-\xi^2}{\xi^2+\eta^2}}\right)\right].
\end{equation}
For such value of $y^*$, we have $|\tilde{q}_1(y^*,t)|=\infty$. 

It appears that there are three distinguished cases $|\xi|>|\eta|$, $|\xi|=|\eta|$, and $|\xi|<|\eta|$, according to the number of singularities of $\tilde{q}_1$. For these three cases, $\epsilon(x,t)$ behaves differently. We express \eqref{epsilon} as
\begin{equation} \label{eps-eq}
\tanh^{-1}\left(\frac{\xi^2+\eta^2}{\eta}\epsilon(x,t)-1\right)-2\eta \epsilon(x,t)=2\eta(x-4\xi t).
\end{equation}
The range of $\epsilon(x,t)$ can be read off from the above equation \eqref{eps-eq}, that is,
$$0\leq \epsilon(x,t) \leq \frac{2\eta}{\xi^2+\eta^2} \quad \mbox{if}\; \eta>0,$$
$$ \frac{2\eta}{\xi^2+\eta^2}\leq \epsilon(x,t) \leq 0 \quad \mbox{if}\; \eta<0.$$
For convenience, we use the notation $\tilde{\epsilon}+1=\frac{\xi^2+\eta^2}{\eta}\epsilon$ in \eqref{eps-eq} to get
\begin{equation}\label{eps-eq-2}
\tanh^{-1}\left(\tilde{\epsilon} \right)-\frac{2\eta^2}{\xi^2+\eta^2} \tilde{\epsilon}=2\eta(x-4\xi t)+\frac{2\eta^2}{\xi^2+\eta^2}, \quad -1\leq \tilde{\epsilon}\leq 1.
\end{equation}
Since $\left.\frac{d \tanh^{-1}(\tilde{\epsilon})}{d\tilde{\epsilon}}\right|_{\tilde{\epsilon}=0}=1$,  the slope of the function 
$$\tanh^{-1}\left(\tilde{\epsilon} \right)-\frac{2\eta^2}{\xi^2+\eta^2} \tilde{\epsilon}$$
at $\tilde{\epsilon}=0$ is positive when $|\xi|>|\eta|$, zero when $|\xi|=|\eta|$, and negative when $|\xi|<|\eta|$. 
The behaviors of $\epsilon(x,t)$ differs with respect to these three cases. 

\begin{figure}[htbp] 
   \centering
   \includegraphics[width=2.5in]{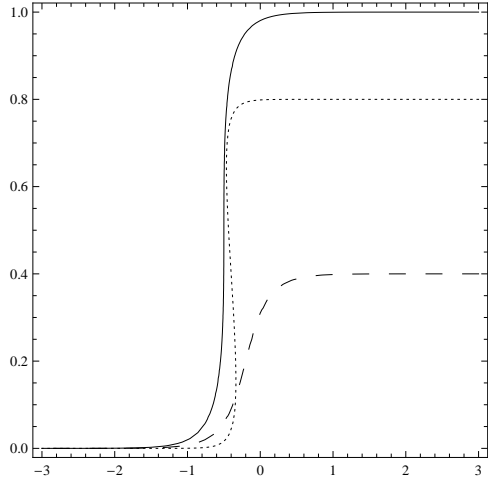} 
   \caption{$\epsilon(x,t) $ v.s. $x$: the large dashed curve corresponds to the case $(\xi,\eta)=(2,1)$, the solid curve corresponds to $(\xi,\eta)=(1,1)$, and the small dashed curve corresponds to $(\xi,\eta)=(1,2)$.}
   \label{fig1}
\end{figure}

The above Figure shows three different curves, as expected. The case $(\xi,\eta)=(2,1)$ gives $\epsilon(x,t)$ whose slope is finite for all $x\in \RR$. The case $(\xi,\eta)=(1,1)$ gives $\epsilon(x,t)$ whose slope is vertical at a point. The case $(\xi,\eta)=(1,2)$ gives $\epsilon(x,t)$ which is multi-valued on some interval of $x$.

 In the following, we give soliton solutions for three different cases. 
\newpage
\subsection{the cases $|\xi|>|\eta|$ and $|\xi|=|\eta|$ (smooth soliton and bursting soliton)} 
\label{smooth-bursting}

The corresponding soliton solutions with $\epsilon(x,t)$ in Figure \ref{fig1} for $(\xi,\eta)=(2,1)$ and $(1,1)$ are shown in the following.

\begin{figure}[htbp] 
   \centering
   \includegraphics[width=2.5in]{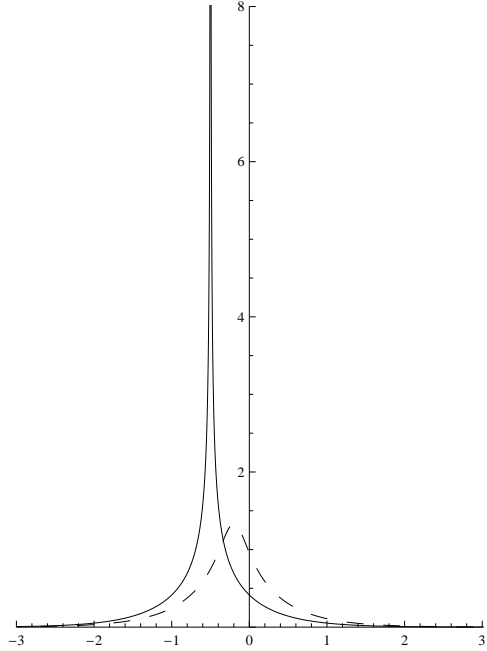} 
   \caption{$|q_1(x,0)| $ v.s. $x$: the solid curve corresponds to $(\xi,\eta)=(1,1)$ and the dashed curve corresponds to  $(\xi,\eta)=(2,1)$.}
   \label{fig2}
\end{figure}

The maximum hight of the bursting soliton above is infinity. We emphasize that the bursting point corresponds to the value of $x$ for $\epsilon(x,t)$ such that the slope of $\epsilon(x,t)$ is vertical. This is easily seen from  
\begin{equation}\label{eps-deriv}
\frac{d\epsilon}{dx}=\frac{2\eta^2}{\xi^2+\eta^2}\frac{1}{D(y,t)}, \quad y=x+\epsilon(x,t),
\end{equation}
where $D(y,t)$ is exactly the denominator of the one soliton $q_1(x,t)$.
This implies that the shift $y=x+\epsilon(x,t)$ by $\epsilon(x,t)$ is very fast in the neighborhood of the singularity of the function $q_1$ for the case $|\xi|=|\eta|$. The type of singularity of $|\tilde{q}_1(y,t)|$ is an essential singularity, but because of the shift by $\epsilon(x,t)$, $|q_1(x,t)|$ is in fact integrable. It may be of our interest to check the first conserved quantity: 
\begin{align*}
\int_{\RR}(\sqrt{|q(x,t)|^2+1}-1)dx &=\int_{\RR}\frac{1}{\cosh^2(2\eta(x-4\xi t+\epsilon(x,t)))-\frac{2\eta^2}{\xi^2+\eta^2}}dx \\
&=\frac{1}{2|\eta|}\int_{\RR}\sech^2(x')dx'\\
&=\frac{1}{|\eta|},
\end{align*}
where in the second inequality we used the change of variable $x-4\xi t +\epsilon(x,t)=x'$ since we know $dx=\frac{dx'}{1+\frac{d\epsilon}{dx}}=\frac{\cosh^2(2\eta x')-\frac{2\eta^2}{\xi^2+\eta^2}}{2|\eta| \cosh^2(2\eta x')}dx'$ from \eqref{eps-deriv}. We see that the first conserved quantity for one soliton solution is free of $\xi$.

\subsection{The case $|\xi|<|\eta|$} 
From equation \eqref{zero-y}, it can be seen that there are two bursting points, and from Figure \ref{fig1}, $\epsilon(x,t)$ is multi-valued for some interval of $x$. Writing $x$ in terms of the $\epsilon$ variable in \eqref{one-soliton} by using \eqref{eps-eq}, we find that the resulting function $q_1(x(\epsilon),t)$ is a single-value function for $\epsilon \in (0, \frac{2\eta}{\xi^2+\eta^2})$ for $\eta>0$, or  $\epsilon \in (\frac{2\eta}{\xi^2+\eta^2},0)$ for $\eta<0$. The following Figure is produced by varying the value of $\epsilon$ $\in (0, \frac{2\eta}{\xi^2+\eta^2})$ and by plotting the corresponding value $(x(\epsilon),q(x(\epsilon),t))$. 
 
We choose $(\xi,\eta)=(\frac{1}{10},\frac{2}{10})$ for a clearer picture. 

\begin{figure}[htbp] 
   \centering
   \includegraphics[width=3.5in]{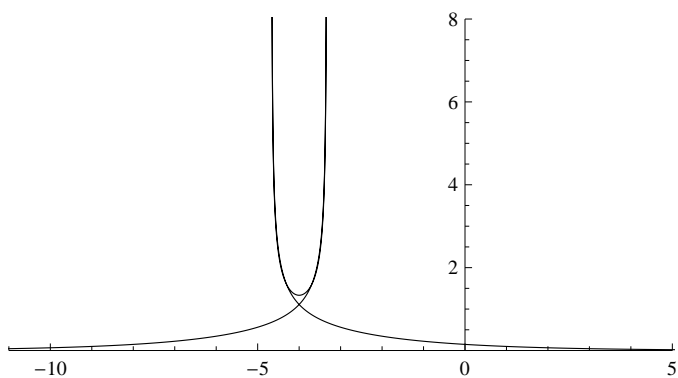} 
   \caption{$|q_1(x,0)|$ v.s. $x$: $(\xi,\eta)=(\frac{1}{10},\frac{2}{10})$.}
   \label{fig3}
\end{figure}

The solution presented above may be an analogue to loop soliton, except that there are two points that meet at infinity, i.e.,
\begin{figure}[htbp]
\centering
\begin{tikzpicture}
\draw[black]  plot [smooth,tension=1] coordinates { (-2,0) (0,0.5) (0.5,1) (0,0.7) (-0.5,1) (0,0.5) (2,0)};

\draw[fill] (0.5,1) circle [radius = 0.05];
\node[above] at (0.5,1) {$\infty$};

\draw[fill] (-0.5,1) circle [radius = 0.05];
\node[above] at (-0.5,1) {$\infty$};

\end{tikzpicture}
\end{figure}

We just mention here that when the ratio $\frac{|\xi|}{|\eta|}$ is small enough, the numerator of $|\tilde{q}_1(y,t)|$ in \eqref{q1-mag} becomes small at $y=4\xi t$. This lowers the U-curve in Figure \ref{fig3} down. As a result, we have the following type of solution. When $\frac{|\xi|}{|\eta|}$ is small enough,

\begin{figure}[htbp]
\centering
\begin{tikzpicture}
\draw[black]  plot [smooth,tension=0.5] coordinates { (-2,0) (0,0.5) (0.5,1) (0,0.1) (-0.5,1) (0,0.5) (2,0)};

\draw[fill] (0.5,1) circle [radius = 0.05];
\node[above] at (0.5,1) {$\infty$};

\draw[fill] (-0.5,1) circle [radius = 0.05];
\node[above] at (-0.5,1) {$\infty$};

\end{tikzpicture}
\end{figure}

\section{Two soliton solution $N=2$} \label{two-soliton-example}
In Subsection \ref{smooth-bursting}, we considered two types of one soliton solutions that are smooth and bursting solitons. In this Section, we focus solely on two soliton solution with the following chosen parameters for $q_2(x,t)$ in \eqref{q-N}, 
$$
\lambda_1=1+i, \quad \lambda_2=2+i
$$
and 
\begin{equation} \label{beta}
\beta_1=a_1|\lambda_1^{-1}-\overline{\lambda}_1^{-1}|, \quad \beta_2=a_2|\lambda_2^{-1}-\overline{\lambda}_2^{-1}|, 
\end{equation}
where
we will fix values of $a_1$ and $a_2$ later on. In Subsection \ref{smooth-bursting}, $\lambda_1$ corresponds to a bursting soliton and $\lambda_2$ corresponds to a smooth soliton if $N=1$.  

\subsection{the case of $a_1=a_2=1$ (Two soliton solution with smooth and bursting peaks) } We first consider the case of $a_1=a_2=1$ in \eqref{beta}. In this case, we simply find that two soliton $q_2(x,t)$ found in \eqref{q-N} consists of a bursting soliton and a smooth soliton as shown in the following Figure, showing that the small peak bypasses the bursting peak as time increases. 
\begin{figure}[htbp] 
   \centering
   \includegraphics[width=3.5in]{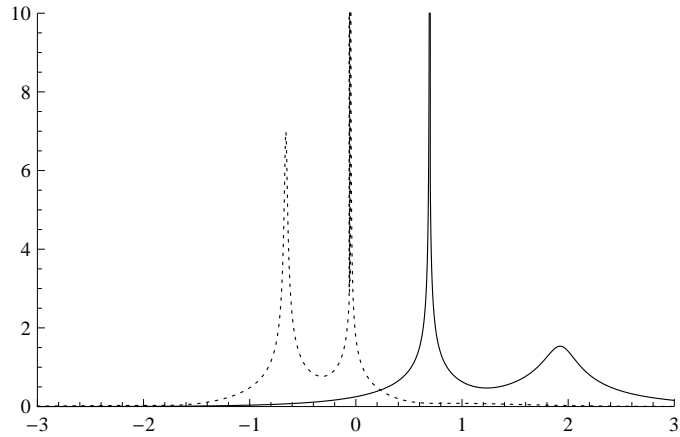} 
   \caption{$|q_2(x,t)| $ v.s. $x$: the dashed curve corresponds to $t=0$ and the solid curve corresponds to $t=\frac{3}{10}$.}
   \label{figTwosoliton}
\end{figure}

The distance between two peaks increases as time increases, since the smaller peak moves faster.


\newpage
\subsection{the case of $a_1=1$ and $a_2=\frac{1}{3}$ (Blowup soliton solution)} \label{blowup-section} Next, we consider the case of $a_1=1$ and $a_2=\frac{1}{3}$ in \eqref{beta}. The corresponding soliton solution $q_2(x,t)$ is found in \eqref{q-N}. 
\begin{figure}[htbp] 
   \centering
   \includegraphics[width=3.5in]{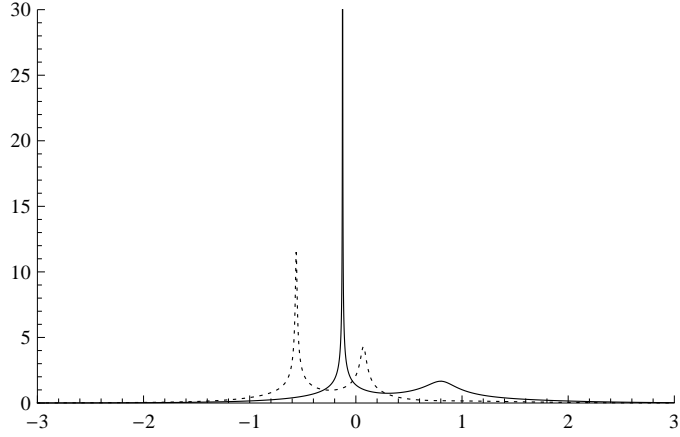} 
   \caption{$|q_2(x,t)| $ v.s. $x$: the dashed curve corresponds to $t=0$ and the solid curve corresponds to $t=\frac{1}{10}$.}
   \label{fig6}
\end{figure}

We observe that the highest peak of $|q_2(x,t)|$ at $t=0$ increases rapidly as time increases, as shown in the case of $t=1/10$. In what follows, we will show that \emph{the above solution blows up in a finite time.}


In order to show blowup, we are concerned with the denominator of $|\tilde{q}_2(y,t)|$ in \eqref{N-soliton-y}, i.e.,
$$D_2(y,t):=1-|\partial_yg(y,t)|^2.$$
Clearly, if $D_2(y,t)=0$, then $|\tilde{q}_2(y,t)|=\infty$. If $|\tilde{q}_2(y,t)|$ is infinity, then there must be a point $x$ such that $|q_2(x,t)|$ is infinity as well. We shall focus on the quantity $D_2(y,t)$.
  
The explicit expression of $D_2(y,t)$ is lengthy, but we shall record it here, 
$$D_2(y,t)=\left(\frac{K_1(y,t)}{K_2(y,t)}\right)^2$$
with 
(denote $p=e^{4t}$ and $q=e^{4y}$)
\begin{align*}
K_1(y,t)&=
3^45^6p^{24}+3^45^6p^{16}q^2-2\cdot3^45^6p^{20}q\\
&\quad  +(-2^33^65^3p^{18}q+2^33^45^319p^{14}q^2)\cos(12t-2x)\\
&\quad \quad \quad -(2^33^35^3p^{14}q^2+2^33^45^313p^{18}q)\sin(12t-2x)\\
&\quad +2^23^25^3107p^{12}q^2+2\cdot3^25^5p^{16}q+2\cdot3^25^3p^8q^3\\
&\quad\quad \quad-2^93^37p^{12}q^2\sin(24t-4x)+2^53^217\cdot31p^{12}q^2\cos(24t-4x)\\
&\quad-2\cdot5^4p^4q^3+(-2^33\cdot5^2149p^{10}q^2+2^33^279p^6q^3)\sin(12t-2x)\\ 
&\quad\quad \quad+(-2^33^3p^{6}q^{3}+2^33^65^2p^{10}q^2)\cos(12t-2x)\\
&\quad+5^2q^4+5^6p^8q^2,\\
\end{align*}
$$K_2(y,t)=\left\{3^25^3p^{12}+5q^2+3^25^3p^8q+2^63^2p^6q\cos(12t-2x)-3\cdot2^37p^6q\sin(12t-2x)+5^3p^4q\right\}^2.$$

We shall study it along the certain characteristic coordinate. The suitable choice comes from the observation that the first three terms in $K_1(y,t)$ cancel to zero if and only if $x=4t$. Clearly, $K_2(4t,t)$ grows faster than $K_1(4t,t)$. Therefore, we find that 
$$\frac{K_1(4t,t)}{K_2(4t,t)} \rightarrow 0, \quad t\rightarrow \infty.$$
While the above does not still imply the finite time blowup, it implies that there exists an infinite peak along the characteristic at $t=\infty$. 

In order to conclude a finite-time blowup, we show the following Figure.
\begin{figure}[htbp] 
   \centering
   \includegraphics[width=3in]{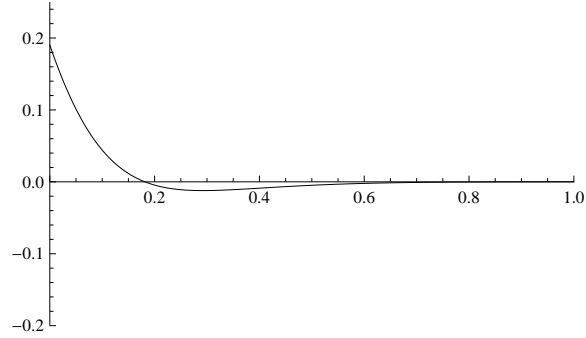} 
   \caption{$\frac{K_1(4t,t)}{K_2(4t,t)}$ v.s. $t$ }
   \label{figblowup}
\end{figure}

We find that the first zero of $\frac{K_1(4t,t)}{K_2(4t,t)}=0$ occurs at $t=t^*=0.18149\cdots.$ This implies that $t^*$ is the time when the infinite peak intersects with the characteristic line $t=\frac{x}{4}$ on the $t$-$x$ coordinate. While the time $t^*$ is not likely to be the minimal blowup time, it still implies the existence of finite time blowup.



\begin{thebibliography}{10}
\bibitem{Ablowitz-Prinari-Trubatch-2004} 
M. J. Ablowitz, B. Prinari, and A. D. Trubatch, \emph{Discrete and Continuous Nonlinear Schr\"{o}dinger systems}, Cambridge University Press, Cambridge, 2004.

\bibitem{Boiti-Gerdjikov-Pempinelli-1986} M. Boiti, V. S. Gerdjikov, and F. Pempinelli,
\emph{The WKIS  System: B\"{a}cklund Transformations, Generalized Fourier Transforms and All That},
Prog. Theor. Phys., {\bf 75} (1986), 1111--1140.

\bibitem{Feng-Inoguchi-Kajiwara-Maruno-2011} B. Feng, J. Inoguchi, K. Kajiwara, K. Maruno, and Y. Ohta,
\emph{Discrete Integrable Systems and Hodograph Transformations Arising from Motions of Discrete Plane Curves},
J. Phys. A: Mathematical and Theoretical, {\bf 44} (2011), 19pp.

\bibitem{Feng-Maruno-Ohta-2014} B. Feng, K. Maruno, and Y. Ohta,
\emph{Self-adaptive moving mesh schemes for short pulse type equations and their Lax pairs},
Pacific Journal of Mathematics for Industry, {\bf 6} (2014).

\bibitem{Matsuno-2007} Y. Matsuno, 
\emph{Multiloop Soliton and Multibreather Solutions of the Short Pulse Model Equation},
J. Phys. Soc. Jpn., {\bf 76} (2007).

\bibitem{Ishimori-1982} Y. Ishimori, 
\emph{A Relationship between the Ablowitz-Kaup-Newell-Segur and Wadati-Konno-Ichikawa Schemes of the Inverse Scattering Method},
J. Phys. Soc. Jpn., {\bf 51} (1982), 3036--3041.

\bibitem{Kundu-1987} A. Kundu, 
\emph{Explicit auto-B\"{a}cklund relation through gauge transformation},
J. Phys. A: Math. Gen., {\bf 20} (1987), 1107--1114.

\bibitem{Levi-Ragnisco-Sym-1984} D. Levi, O. Ragnisco, and A. Sym,
\emph{The B\"{a}cklund transformations for nonlinear evolution equations which exhibit exotic solitons},
Phys. Lett. A, {\bf 100} (1984), 7--10.

\bibitem{Liu-Pelinovsky-Sakovich-2009} Y. Liu, D. Pelinovsky, and A. Sakovich, 
\emph{Wave breaking in the short-pulse equation},
Dynam. Part. Differ. Eq., {\bf 6} (2009), 291--310.

\bibitem{Sakovich-2006}  A. Sakovich and S. Sakovich,
\emph{Solitary wave solutions of the short pulse equation},
J. Phys. A: Mathematical and General, {\bf 39} (2006), 361--367.

\bibitem{Shimabukuro-present} Y. Shimabukuro, 
\emph{Global solution of the Wadati-Konno-Ichikawa equation with small Initial data}, preprint, (2016), arXiv:1612.07579. 

\bibitem{Shimizu-Wadati-1980} T. Shimizu and M. Wadati, 
\emph{A New Integrable Nonlinear Evolution Equation}, 
Prog. Theor. Phys., {\bf 63} (1980), 808--820.

\bibitem{Svistunov-1995} B. V. Svistunov, 
\emph{Superfluid turbulence in the low-temperature limit},
Phys. Rev. B, {\bf 52} (1995), 3647--3653.

\bibitem{Gorder-2016} R. A. Van Gorder,
\emph{Solitons and nonlinear waves along quantum vortex filaments under the low-temperature two-dimensional local induction approximation},
Phys. Rev. E, {\bf 93} (2016), pp 15.

\bibitem{Wadati-Konno-Ichikawa-1979} M. Wadati, K. Konno, and Y. Ichikawa,
\emph{New Integrable Nonlinear Evolution Equations}, 
Journal of the Physical Society of Japan, {\bf 47} (1979), 1698--1700.

\bibitem{Wadati-Sogo-1983} M. Wadati and M. Sogo,
\emph{Gauge Transformations in Soliton Theory},
J. Phys. Soc. Jpn., {\bf 52} (1983), 394--338. 

\bibitem{Zakharov-Takhtadzhyan-1978} V. E. Zakharov and L. A. Takhtadzhyan, 
\emph{Equivalence of the nonlinear Schr\"{o}dinger equation and the equation of a Heisenberg Ferromagnet},
TMF, {\bf 38} (1979), 26--35. 

\bibitem{Zakharov-Mikhailov-1978} V. E. Zakharov and A. V. Mikhailov,
\emph{Relativistically invariant two-dimensional models of field theory which are 
integrable by means of the inverse scattering problem method},
Sov. Phys. JETP, {\bf 74} (1978), 1953--1973.



\end{thebibliography}
\end{document}